\documentclass{ieeeconf}
\usepackage{graphicx} % include this line if your document contains figures

\usepackage{amsmath} % math package
\usepackage{amssymb}
\usepackage{bm} 

\usepackage{tikz}
\usetikzlibrary{calc,patterns,decorations.pathmorphing,decorations.markings,decorations.pathreplacing}

\usepackage{textcomp}
\usepackage{hyperref}
\usepackage{lipsum}

\newcommand\copyrighttext{
	\footnotesize \textcopyright 2021 the
	authors. This work has been accepted to IFAC for publication under a Creative Commons	Licence CC-BY-NC-ND.\\ {DOI: \href{https://doi.org/10.1016/j.ifacol.2021.08.550}{10.1016/j.ifacol.2021.08.550}} }
\newcommand\copyrightnotice{
	\begin{tikzpicture}[remember picture,overlay]
		\node[anchor=south,yshift=10pt] at (current page.south) {\fbox{\parbox{\dimexpr\textwidth-\fboxsep-\fboxrule\relax}{\copyrighttext}}};
	\end{tikzpicture}%
}

\PassOptionsToPackage{dvipsnames}{xcolor}
\renewcommand{\theenumi}{\roman{enumi}}%
\newcommand{\nat}{\mathbb{N}}

\newtheorem{thm}{Theorem}
\newtheorem{defn}[thm]{Definition}
\newtheorem{assum}[thm]{Assumption}
\newtheorem{rem}[thm]{Remark}

\title{\LARGE \bf Constraint-adaptive MPC for large-scale systems: \\ Satisfying state constraints without imposing them}

\author{S.A.N. Nouwens$^{1}$, B. de Jager$^{1}$, M.M. Paulides$^{2,3}$, W.P.M.H. Heemels$^{1}$% <-this % stops a space
	\thanks{This research is supported by KWF Kankerbestrijding and NWO Domain AES, as part of their joint strategic research programme: Technology for Oncology II. The collaboration project is co-funded by the PPP Allowance made available by Health$\sim$Holland, Top Sector Life Sciences \& Health, to stimulate public-private partnerships.}% <-this % stops a space
	\thanks{$^{1}$Control Systems Technology, Department of Mechanical Engineering, Eindhoven University of Technology, Eindhoven, The Netherlands
	}%
	\thanks{$^{2}$Electromagnetics for Care \& Cure, Department of Electrical Engineering, Eindhoven University of Technology, Eindhoven, The Netherlands}%
	\thanks{$^{3}$Department of	Radiotherapy, Erasmus University Medical Center Cancer Institute, Rotterdam, The Netherlands}%
}

\begin{document}
\bstctlcite{IEEEexample:BSTcontrol} %to force et al.

\maketitle
\thispagestyle{empty}
\pagestyle{empty}

\copyrightnotice
\begin{abstract} % Abstract of not more than 250 words.
	Model Predictive Control (MPC) is a successful control methodology, which is applied to increasingly complex systems. However, real-time feasibility of MPC can be challenging for complex systems, certainly when an (extremely) large number of constraints have to be adhered to. For such scenarios with a large number of state constraints, this paper proposes two novel MPC schemes for general nonlinear systems, which we call constraint-adaptive MPC. These novel schemes dynamically select at each time step a (varying) set of constraints that are included in the on-line optimization problem. Carefully selecting the included constraints can significantly reduce, as we will demonstrate, the computational complexity with often only a slight impact on the closed-loop performance. Although not all (state) constraints are imposed in the on-line optimization, the schemes still guarantee recursive feasibility and constraint satisfaction. A numerical case study illustrates the proposed MPC schemes and demonstrates the achieved computation time improvements exceeding two orders of magnitude without loss of performance.
\end{abstract}
%\begin{keyword}
%Model Predictive Control; Large-Scale Systems; Adaptive Constraints.
%\end{keyword}

%===============================================================================

\section{Introduction}
\subsection{Motivation}
Model Predictive Control (MPC) is an advanced feedback control strategy with widespread industrial application \cite{Mayne2014,Dotoli2015}. While MPC exists in numerous forms tailored to the specific application, its basic structure remains the same. Indeed, in general, MPC is an optimization-based controller that minimizes a finite-horizon performance criterion subject to system dynamics and constraints over a future sequence of control inputs. At each point of time this optimization problem is solved based on the (estimated) state of the controlled plant. By selecting the first value of the optimal control sequence as the control input applied to the plant and using a receding-horizon principle to obtain the control input, the MPC feedback policy is created.

As mentioned, the control input in (implicit) MPC is obtained by solving an on-line optimization problem at each time step. As such, the real-time feasibility of MPC is determined to a large extent by the computational complexity of this optimization problem (and, of course, the available computational hardware). Certainly for complex systems with a high number of decision variables and constraints, the computational complexity of the optimization problem can prohibit real-time implementation. Therefore, numerous methods have been proposed that aim to reduce the computational complexity, such as explicit MPC, model reduction, and numerical solver optimizations, see, e.g., \cite{Rawlings2019,Hovland2006,Jerez2011}. 

The computational complexity of MPC schemes depends on, among others, the number of decision variables, the model dimension, and the number of constraints \cite{Boyd2009}. In this paper, we are particularly interested in systems with an (extremely) large number of inequality (state) constraints, which can lead to computational intractability for real-time MPC. Dynamical systems with many state constraints are encountered in a diverse range of scenarios, including the case of controlling partial differential equations (PDEs). Certainly, in these cases MPC control is challenging \cite{Altmuller2012}. Typically, the spatial discretization of the PDE results in a large-scale highly complex model. Furthermore, if there are constraints defined on the continuous spatial domain, spatial discretization will yield many constraints. This problem is, for instance, encountered in Magnetic Resonance (MR) guided hyperthermia in cancer treatments \cite{Hendrikx2018,Paulides2013}. Here, a temperature upper bound on the patients' tissue results in a large number of inequality constraints on the discrete spatial domain. As a result, both the large model and the high number of constraints complicate the real-time implementation of MPC strategies for PDE-based systems. While model reduction techniques for large-scale systems are available, they do not reduce the number of inequality constraints \cite{Antoulas2000}, and, hence, building MPC schemes for such applications remains a true challenge. This observation forms a strong motivation for the present work. 

\subsection{State-of-the-art}
For systems with many inequality constraints, a natural idea is to reduce the number of inequality constraints to obtain real-time feasibility of the MPC scheme. For example, it is possible to identify and remove redundant constraints prior to solving the optimization problem. However, generally, this is intractable in cases with a large number of inequality constraints \cite{Paulraj2010}. Alternatively, one could aim for online constraint removal as proposed by \cite{Jost2013}. Unfortunately, this approach requires extensive offline computations that do not scale well to a (very) large number of constraints. Other recent techniques use machine learning to either approximate the MPC control law \cite{Hertneck2018} or use a hybrid approach where machine learning is used to better warm start an active set algorithm \cite{Klauco2019}. While these methods showed clear improvements in computation time, they require extensive offline training.

\subsection{Contributions}
As solving the MPC problem within the sample time for systems with a large number of inequality constraints remains to be impossible in many situations, novel MPC strategies are needed that ensure constraint satisfaction and good performance properties while being computationally tractable. In this research we propose two novel real-time feasible MPC schemes for models with many inequality constraints. The key idea behind our schemes is based on adaptively including only a (small) well-chosen and state-dependent subset of the state constraints in the on-line optimization problem, which then becomes tractable to solve. For this reason, we call our new schemes {\em constraint-adaptive MPC} (ca-MPC). Critically, our approach guarantees constraint satisfaction and recursive feasibility under appropriate conditions, as we will prove. So, although not all (state) constraints are imposed in the on-line optimization, we will still be able to prove that all constraints are satisfied under the proposed MPC laws.

The remainder of this paper is structured as follows. First, we introduce a regular MPC setup and the problem formulation in Section~\ref{sec:setup_and_problem}. In Section~\ref{sec:invariant}, we treat the case where the state constraint set is controlled invariant, propose the corresponding ca-MPC approach and establish recursive feasibility and constraint satisfaction. In Section~\ref{sec:terminal}, we consider the case where the state constraint set is not controlled invariant by including a suitably chosen terminal set. In Section~\ref{sec:sim_results}, a simulation study is performed for a double integrator system to demonstrate the benefits of our new method. There we will observe a significant reduction in computation time (two orders of magnitude) without loss of performance. Last, in Section~\ref{sec:conclusion} we present the conclusions and recommendations.

%%%%%%%%%%%%%%%%%%%%%%%%%%%%%%%%

\section{MPC setup and problem formulation}\label{sec:setup_and_problem}
In this section, we introduce the system to be controlled, the MPC setup, and the problem formulation. 

\subsection{System description} 
In this paper, we consider plants that are described by the discrete-time time-invariant nonlinear system
\begin{align}\label{eq:system_description}
	&\bm{x}_{k+1} = f(\bm{x}_k,\bm{u}_k).
\end{align}
Here, $\bm{x}_k\in\mathbb{R}^n$ and $\bm{u}_k\in\mathbb{R}^m$ denote the plant states and the inputs, respectively at discrete time $k\in\mathbb{N}$. Furthermore, $f:\mathbb{R}^n\times\mathbb{R}^m\rightarrow\mathbb{R}^n$ denotes a possibly nonlinear state transition function. The system \eqref{eq:system_description} is subject to state and input constraints that are given by
\begin{align} \label{eq:state_and_input_constraints}
	\bm{x}_k \in \mathbb{X},\quad	\bm{u}_k \in \mathbb{U},\quad k\in\mathbb{N},
\end{align}
where
\begin{subequations}
  \begin{align}
	&\mathbb{X} := \{\bm{x}\in\mathbb{R}^n\mid \bm{Cx}\leq \bm{b}\},\\
	&\mathbb{U} := \{\bm{u}\in\mathbb{R}^m\mid \bm{C}_u\bm{u}\leq \bm{b}_u\},
\end{align}  
\end{subequations}
are assumed to be non-empty sets with $\bm{C}\in\mathbb{R}^{n_x\times n}$ and $\bm{C}_u\in\mathbb{R}^{n_u\times m}$. Here, $n_x$ and $n_u$ denote the number of state and input constraints, respectively.

\subsection{MPC setup} 
Based on system \eqref{eq:system_description} and constraints \eqref{eq:state_and_input_constraints}, a MPC setup given state $\bm{x}_k$ at time $k\in\mathbb{N}$ can be formulated as 
\begin{subequations}\label{eq:basic_mpc}
	\begin{align}
		\underset{\mathcal{U}_k:=(\bm{u}_{0|k},\dots,\bm{u}_{N-1|k})}{\text{minimize}}\qquad&\ell_T(\bm{x}_{N|k}) + \sum_{i=0}^{N-1}\ell(\bm{x}_{i|k},\bm{u}_{i|k}),\label{eq:basic_mpc_a}\\
		\text{subject to    }\qquad&\bm{x}_{i+1|k} = f(\bm{x}_{i|k},\bm{u}_{i|k}),\label{eq:basic_mpc_b}\\
		&\bm{x}_{0|k} = \bm{x}_k, \label{eq:basic_mpc_c}\\
		&\bm{x}_{i|k} \in \mathbb{X},\quad i = 1,\dots,N,\label{eq:basic_mpc_d}\\
		&\bm{u}_{i|k} \in \mathbb{U},\quad i = 0,\dots,N-1.\label{eq:basic_mpc_e}
	\end{align}
\end{subequations}
Here, $\ell(\cdot)$, $\ell_T(\cdot)$, $\bm{x}_{i|k}$, and $\bm{u}_{i|k}$ denote the stage cost, the terminal cost, the predicted state, and the predicted input, respectively. In particular, the $i|k$ subscript is used to denote the $i$-th prediction at time $k$. For example, $\bm{x}_{i|k}$ denotes the prediction of $\bm{x}_{k+i}$ made at time $k$. Additionally, the stage and terminal cost quantify the performance in terms of the predicted states $\bm{x}_{i|k}$ and inputs $\bm{u}_{i|k}$, $i=0,1,2,\ldots,N$. For the optimization problem \eqref{eq:basic_mpc} we denote the set of feasible inputs for state $\bm{x}_k$ by 
\begin{align}
\mathbb{U}_F(\bm{x}_k) :=\{ (\bm{u}_{0|k},\dots,\bm{u}_{N-1|k}) \in \mathbb{U}^N \mid \eqref{eq:basic_mpc_b}-\eqref{eq:basic_mpc_e}\}.
\end{align}
Moreover, we denote the set of feasible states of \eqref{eq:basic_mpc} by 
\begin{align} \label{eq:feasible}
\mathbb{X}_F :=\{ \bm{x} \in \mathbb{X} \mid \mathbb{U}_F(\bm{x}) \neq \emptyset \}.
\end{align}
Lastly, under suitable assumptions on $\ell(\cdot)$, $\ell_T(\cdot)$, $\mathbb{X}$, and $\mathbb{U}$, a minimizer of this optimization problem exists for all $\bm{x}_k\in \mathbb{X}_F$ and we denote by $\mathcal{U}^\star_{k} := (\bm{u}^\star_{0|k},\dots,\bm{u}^\star_{N-1|k})$, a particular one at time $k\in\mathbb{N}$, i.e., 
\begin{align}
\mathcal{U}^\star_{k} \in \arg \min_{\mathcal{U}_k \in \mathbb{U}_F(\bm{x}_k) } \ell_T(\bm{x}_{N|k}) + \sum_{i=0}^{N-1}\ell(\bm{x}_{i|k},\bm{u}_{i|k}).
\end{align}
The optimal predicted state sequence corresponding to $\mathcal{U}^\star_{k}$ is denoted by $\mathcal{X}_k^\star = (\bm{x}_{1|k}^\star,\dots,\bm{x}_{N|k}^\star)$. 

Using a receding horizon implementation, the MPC problem \eqref{eq:basic_mpc} is turned into a feedback law by applying the first computed input in $\mathcal{U}_k^\star$ on the real plant \eqref{eq:system_description}, i.e., \begin{equation}
\bm{u}_k := K_{MPC}(\bm{x}_k):= \bm{u}^\star_{0|k}. \label{eq:MPC} \end{equation}

\subsection{Problem statement}

As discussed in the introduction, to enable real-time feasibility of the MPC scheme \eqref{eq:MPC}, we need to be able to compute a minimizer to the optimization problem \eqref{eq:basic_mpc} within the sampling period that corresponds to the discrete-time plant \eqref{eq:system_description}. This becomes particularly challenging if the number of constraints, $n_x$, in the state constraint set $\mathbb{X}$ becomes very large. The goal of this paper is to provide novel MPC schemes that are computationally more tractable by replacing the state constraint set in the online optimization problem by one with a significantly lower number of (state) constraints, while still leading to good performance properties and guaranteeing constraint satisfaction \eqref{eq:state_and_input_constraints}. 

%%%%%%%%%%%%%%%%%%%%%%%%%%%%%%%%

\section{ca-MPC under controlled invariance of original constraint set} \label{sec:invariant} 

As already hinted upon in the previous section, we will replace the constraint set $\mathbb{X}$ in the on-line optimization problem \eqref{eq:basic_mpc} by a ``simpler'' one that contains a small subset of the constraints in $\mathbb{X}$, thereby making the on-line optimization problem computationally tractable. Regardless, constraint satisfaction must be assured for all times. In particular, we are going to replace $\mathbb{X}$ by a well-crafted combination of a reduced constraint set
\begin{equation}\label{eq:construction_red_constraint_set}
	\mathbb{X}_r(\bm{x}_k) := \{\bm{x}\in\mathbb{R}^n | \bm{C}_j\bm{x} \leq \bm{b}_j, \quad j \in \mathcal{J}(\bm{x}_k)\},
\end{equation}
where $\mathcal{J}(\bm{x}_k) \subset \mathbb{N}_{[1,n_x]}$ is the index set that captures the selected constraints, and a so-called delta set $\Delta(\bm{x}_k)$ based on the current state $\bm{x}_k$ (and possibly other information available at time $k\in\mathbb{N}$). The exact construction will be explained in detail below. In this section, we first consider the case where $\mathbb{X}$ is controlled invariant.

\begin{defn}
The set $\mathbb{X}$ is controlled invariant for \eqref{eq:system_description} with the input set $\mathbb{U}$, if for all $\bm{x} \in \mathbb{X}$ there exists an $\bm{u}\in\mathbb{U}$ such that $f(\bm{x},\bm{u}) \in \mathbb{X}$.
\end{defn}

The reduced MPC problem at time $k\in\nat$ for state $\bm{x}_k$, where $\mathbb{X}$ is replaced by a combination of $ \mathbb{X}_r(\bm{x}_k)$ and $\Delta(\bm{x}_k)$, is given by
\begin{subequations}\label{eq:reduced_mpc_scheme}
	\begin{align}
		\underset{\bm{u}_{0|k},\dots,\bm{u}_{N-1|k}}{\text{minimize}}\qquad&\ell_T(\bm{x}_{N|k}) + \sum_{i=0}^{N-1}\ell(\bm{x}_{i|k},\bm{u}_{i|k}),\\
		\text{subject to }\qquad&\text{\eqref{eq:basic_mpc_b}, \eqref{eq:basic_mpc_c}, \eqref{eq:basic_mpc_e}},\nonumber\\ 
		&\bm{x}_{1|k} \in \mathbb{X}_r(\bm{x}_k)\cap\Delta(\bm{x}_k),\label{eq:reduced_mpc_scheme_d}\\
		&\bm{x}_{i|k} \in \mathbb{X}_r(\bm{x}_k),\quad i = 2,\dots,N.\label{eq:reduced_mpc_scheme_e}
	\end{align}
\end{subequations}
A few comments are in order. First, only the first predicted state, $\bm{x}_{1|k}$, is constrained by both the reduced constraint set $\mathbb{X}_r(\bm{x}_k)$ and the delta set $\Delta(\bm{x}_k)$, while the other states are constrained only by $\mathbb{X}_r(\bm{x}_k)$. Second, the reduced constraint set depends on $\bm{x}_k$, which leads to the adaptive nature of the set. Last, we define the feasible set for \eqref{eq:reduced_mpc_scheme} as $\mathbb{X}_F^r := \{\bm{x}_k\in\mathbb{X}\mid\mathbb{U}^r_F(\bm{x}_k)\neq\emptyset\}$, where $\mathbb{U}_F^r(\bm{x}_k):=\{\mathcal{U}_k\in\mathbb{U}^{N}\mid\eqref{eq:basic_mpc_b},\eqref{eq:basic_mpc_c},\eqref{eq:basic_mpc_e},\eqref{eq:reduced_mpc_scheme_d},\eqref{eq:reduced_mpc_scheme_e}\}$. Moreover, the corresponding control law is denoted by $K_{MPC}^r: \mathbb{X}_F^r \rightarrow \mathbb{U}$ and defined similarly as in \eqref{eq:MPC}.

We now proceed with proposing a construction for the reduced constraint set $ \mathbb{X}_r: \mathbb{X} \rightrightarrows \mathbb{R}^n$ according to \eqref{eq:construction_red_constraint_set} and guarantee recursive feasibility and constraint satisfaction by providing suitable conditions on $\Delta: \mathbb{X} \rightrightarrows \mathbb{R}^n$. 
 
\begin{figure}[!ht]
	\centering
	\begin{tikzpicture}[join=round]
		\definecolor{red}{HTML}{eb2f06};
		\definecolor{blue}{HTML}{2980b9};
		\definecolor{green}{HTML}{44bd32};
		
		%	Draw constraint set (counter clockwise starting from left vertical)
		\draw[very thick] (0.8, 1.8)--(0.4, 3.8);
		\draw[very thick] (0.3, 2.2)--(4.6, 2.0);
		\draw[very thick] (3.6, 1.85)--(7, 3.5);
		\draw[very thick] (7, 3)--(5.5, 5);
		\draw[very thick] (6.3, 4.7)--(2.7, 5);
		\draw[very thick, color = red] (4.5, 5)--(1, 4.0);
		\draw[very thick, color = red] (3.5, 5.2)--(0.1, 3.2);
		
		%	Draw state trajectory
		\draw[very thick, dashed]  (2.5,3.6) -- (3.4,4.3) -- (5.0,4.6) --  (6.8,4.4) ;
		\draw[fill=black] (2.5,3.6) circle(0.05) node[below] {$\bm{x}_{0|k}$};
		\draw[fill=black] (3.4,4.3) circle(0.05) node[left] {$\bm{x}_{1|k}$};
		\draw[fill=black] (5.0,4.6) circle(0.05) node[below] {$\bm{x}_{2|k}$};
		\draw[fill=black] (6.8,4.4) circle(0.05) node[below] {$\bm{x}_{3|k}$};
		
		%	Draw Delta set
		\draw[very thick, color=green] (2.5,3.5) 	circle(1.38);
		\node[color=green] at (2.3,2.5) {$\Delta(\bm{x}_k)$};
		
		%	Annotate sets
		\node[] at (6,3.5) {$\mathbb{X}$};
		\node[color=red] at (0.75,4.5) {$\mathbb{X}_r(\bm{x}_k)$};
	\end{tikzpicture}
	\caption{Illustrating ca-MPC when $\mathbb{X}$ is controlled invariant. Clearly, $\bm{x}_{1|k}$ \textit{cannot} be outside of $\mathbb{X}$ while $\bm{x}_{2|k},\dots,\bm{x}_{N|k}$ are not necessarily constrained to $\mathbb{X}$ as illustrated by $\bm{x}_{3|k}$.}
	%Observe that $\bm{x}_{1|k}$ \textit{cannot} be outside of $\mathbb{X}$ when constrained to $\mathbb{X}_r(\bm{x}_k) \cap \Delta(\bm{x}_k)$. Last, $\{\bm{x}_{2|k},\dots,\bm{x}_{N|k}\}$ is not necessarily constrained to $\mathbb{X}$ as illustrated by $\bm{x}_{3|k}$.}
	\label{fig:ca-MPC_illustration_CI}
\end{figure}
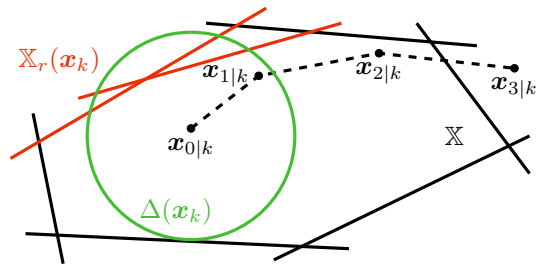

 To select the relevant constraints in \eqref{eq:construction_red_constraint_set} via $\mathcal{J}(\bm{x}_k)$ given a state $\bm{x}_k$, we introduce the concept of a safety set. Here, a safety set $\mathbb{X}_s\subseteq \mathbb{X}$ is defined as a tightened constraint set based on non-negative vector $\bm{\alpha}\geq\bm{0}$, which can be seen as a ``safety margin'' or a ``margin to constraint violation.'' In particular, $\mathbb{X}_s := \{\bm{x}\in\mathbb{R}^n | \bm{C}\bm{x} \leq \bm{b} - \bm{\alpha}\}$. Given $\bm{x}_k$ and $\mathbb{X}_s$, we include constraints in $\mathbb{X}_r(\bm{x}_k)$ based on the constraint violations of the safety set:
\begin{equation} \label{eq:J}
	\mathcal{J}(\bm{x}_k) := \{j \in \mathbb{N}_{[1,n_x]}\mid \bm{C}_j\bm{x}_k > b_j - {\alpha}_{j} \}.
\end{equation}
This ensures that once the state approaches the boundary of $\mathbb{X}$, the relevant constraints are included in $\mathbb{X}_r(\bm{x}_k)$ as, loosely speaking, for these constraints the safety margin has become too small. Observe that the safety set itself is not included in the MPC problem. Furthermore, note that the computation of \eqref{eq:J} and thus computing $\mathbb{X}_r(\bm{x}_k) $ can be performed efficiently and only has to be done once at each time step. We illustrate the MPC setup from \eqref{eq:reduced_mpc_scheme} with the reduced constraint set, delta set, and predicted state trajectory in Fig.~\ref{fig:ca-MPC_illustration_CI}.

While the safety set tries to include relevant constraints in $\mathbb{X}_r(\bm{x}_k)$, it does not guarantee recursive feasibility and constraint satisfaction automatically. To obtain these properties, conditions have to be imposed on $\Delta(\bm{x}_k)$.

\begin{thm}\label{lem:rec_feas_control_invariant}
	Consider the ca-MPC scheme \eqref{eq:reduced_mpc_scheme} with $\mathbb{X}$ controlled invariant for \eqref{eq:system_description} with the input set $\mathbb{U}$. If for all states $\bm{x}\in\mathbb{X}$ and corresponding reduced constraint set $\mathbb{X}_r(\bm{x})$, the delta set $\Delta(\bm{x})$ satisfies the properties
	\begin{enumerate}
		\item $\mathbb{X}_r(\bm{x})\cap\Delta(\bm{x}) \subseteq \mathbb{X}$,\label{cond:delta_cond1}
		\item $\exists \bm{\bar{u}}\in \mathbb{U}$ such that $f(\bm{x},\bm{\bar{u}})\in\mathbb{X}_r(\bm{x})\cap\Delta(\bm{x})$, \label{cond:delta_cond2}		
	\end{enumerate}
	then the feasible set is equal to $\mathbb{X}$, i.e., $\mathbb{X}_F^r = \mathbb{X}$ and \eqref{eq:reduced_mpc_scheme} is recursively feasible in the sense that if $\bm{x} \in \mathbb{X}_F^r$, then $f(\bm{x},K_{MPC}^r(\bm{x})) \in \mathbb{X}_F^r$. Moreover, constraint satisfaction is guaranteed for the closed-loop system
	\begin{equation}
	\bm{x}_{k+1} = f(\bm{x}_k,\bm{u}_k), \ \bm{u}_k= {K}_{MPC}^r(\bm{x}_k) 
	\end{equation}
	in the sense that for all trajectories of this system with $\bm{x}_0 \in \mathbb{X}_F^r = \mathbb{X}$ the constraints in \eqref{eq:state_and_input_constraints} are satisfied. 
\end{thm}
Due to space constraints, all proofs are omitted.

We now have established sufficient conditions on $\Delta: \mathbb{X} \rightrightarrows \mathbb{R}^n$ that guarantee recursive feasibility and constraint satisfaction. However, it is still unclear how to construct a $\Delta(\bm{x})$ that satisfies Conditions~\eqref{cond:delta_cond1}-\eqref{cond:delta_cond2} in Theorem~\ref{lem:rec_feas_control_invariant}. To this end, we propose a construction for $\Delta(\bm{x})$ that satisfies both Conditions~\eqref{cond:delta_cond1} and~\eqref{cond:delta_cond2}. An example of a simple set $\Delta(\bm{x})$ for $\bm{x} \in \mathbb{X}$ given an admissible input $\bm{\bar{u}}$ such that $f(\bm{x},\bm{\bar{u}}) \in \mathbb{X}$ is a sphere: $\Delta(\bm{x}) = \{\bm{\tilde{x}}\in\mathbb{R}^n\mid \|f(\bm{x},\bm{\bar{u}}) - \bm{\tilde{x}}\|_p \leq c(\bm{x})\}$, $p\in \{\mathbb{N},\infty\}$. Here, the radius $c(\bm{x})$ depends on $\bm{x}$ and is chosen such that $\mathbb{X}_r(\bm{x})\cap\Delta(\bm{x}) \subseteq \mathbb{X}$ holds. Note that a $\Delta(\bm{x})$ satisfying the required conditions can always be obtained by selecting $c(\bm{x})=0$ (although that is a very restrictive choice). Depending on the choice of the norm $\|\cdot \|_p$ different state constraint sets are obtained in the on-line MPC optimization problem, e.g., for $p=1,\infty$ we obtain polyhedral sets, while for $p=2$, we obtain quadratic-type constraints as seen in Fig.~\ref{fig:ca-MPC_illustration_CI}.

%%%%%%%%%%%%%%%%%%%%%%%%%%%%%%%%

\section{CA-MPC with terminal sets}\label{sec:terminal}
In the previous section, we assumed that $\mathbb{X}$ was controlled invariant. While this resulted in rather basic conditions regarding recursive feasibility and constraint satisfaction, it is often not the case that the original constraint set $\mathbb{X}$ is controlled invariant, or that there is a natural (sufficiently large) and easy-to-compute controlled invariant set inside $\mathbb{X}$ to replace $\mathbb{X}$. Alternatively, simple terminal sets are more commonplace in guaranteeing recursive feasibility for MPC schemes \cite{Mayne2000}. Therefore, it is of interest to extend the previous results to ca-MPC schemes based on terminal sets, for which we assume the following:
\begin{assum} \label{ass:target_set}
	The set $\mathbb{X}_T\subset \mathbb{X}$ is a nonempty and controlled invariant set for \eqref{eq:system_description} with input set $\mathbb{U}$.
\end{assum}

We start by introducing a standard MPC setup with terminal set as 
\begin{subequations}\label{eq:regular_MPC_terminal}
	\begin{align}
		\underset{\bm{u}_{0|k},\dots,\bm{u}_{N-1|k}}{\text{minimize}}\qquad&\ell_T(\bm{x}_{N|k}) + \sum_{i=0}^{N-1}\ell(\bm{x}_{i|k},\bm{u}_{i|k}),\label{eq:regular_MPC_terminal_a}\\
		\text{subject to }\qquad&\bm{x}_{i+1|k} = f(\bm{x}_{i|k},\bm{u}_{i|k}),\label{eq:regular_MPC_terminal_b}\\
		&\bm{x}_{0|k} = \bm{x}_{1|k-1}^\star=\bm{x}_k,\label{eq:regular_MPC_terminal_c}\\
		&\bm{x}_{i|k} \in \mathbb{X},\quad i = 1,\dots,N-1, \label{eq:regular_MPC_terminal_d}\\
		&\bm{x}_{N|k} \in \mathbb{X}_T,\label{eq:regular_MPC_terminal_e}\\
		&\bm{u}_{i|k} \in \mathbb{U},\quad i = 0,\dots,N-1.\label{eq:regular_MPC_terminal_f}
	\end{align}
\end{subequations}
Observe that the final predicted state $\bm{x}_{N|k}$ is constrained to be contained in the terminal set $\mathbb{X}_T$. 

Now we propose a modification of \eqref{eq:regular_MPC_terminal} to obtain a reduced MPC problem with terminal set at time $k\in\nat$ with state $\bm{x}_k=\bm{x}_{1|k-1}^\star$. This new MPC is parameterized by a state sequence $\mathcal{X}_{k-1}^\star=(\bm{x}^\star_{1|k-1},\ldots,\bm{x}^\star_{N|k-1})$ and given by
\begin{subequations}\label{eq:reduced_MPC_terminal}
	\begin{align}
		\underset{\bm{u}_{0|k},\dots,\bm{u}_{N-1|k}}{\text{minimize}}\qquad&\ell_T(\bm{x}_{N|k}) + \sum_{i=0}^{N-1}\ell(\bm{x}_{i|k},\bm{u}_{i|k}),\label{eq:reduced_MPC_terminal_a}\\
		\text{subject to }\qquad&\text{\eqref{eq:regular_MPC_terminal_b}, \eqref{eq:regular_MPC_terminal_c}, \eqref{eq:regular_MPC_terminal_e}, \eqref{eq:regular_MPC_terminal_f}},\nonumber\\
% 		\qquad&\bm{x}_{i+1|k} = f(\bm{x}_{i|k},\bm{u}_{i|k}),\label{eq:reduced_MPC_terminal_b}\\
% 		&\bm{x}_{0|k} = \bm{x}_{1|k-1}^\star=\bm{x}_k,\label{eq:reduced_MPC_terminal_c}\\ \nonumber
		&\bm{x}_{i|k} \in \mathbb{X}_r(\bm{x}^\star_{i+1|k-1}) \cap \Delta(\bm{x}^\star_{i+1|k-1}), \nonumber\\ 
		&\quad i = 1,\dots,N-1 \label{eq:reduced_MPC_terminal_d}.
% 		&\bm{x}_{N|k} \in \mathbb{X}_T,\label{eq:reduced_MPC_terminal_e}\\
% 		&\bm{u}_{i|k} \in \mathbb{U},\quad i = 0,\dots,N-1.\label{eq:reduced_MPC_terminal_f}
	\end{align}
\end{subequations}
 First of all, let us note that the reduced constraint set $\mathbb{X}_r(\cdot)$ is based on using the optimal predicted state sequence $\mathcal{X}_{k-1}^\star$ at the previous time step (and not only based on $\bm{x}_k$). Second, note that the construction of $\mathbb{X}_r(\cdot)$ can remain unchanged and, hence, is given by \eqref{eq:construction_red_constraint_set} by now exploiting the optimal predicted sequence $\mathcal{X}_{k-1}^\star$. Third, the delta set $\Delta(\bm{x}^\star_{i+1|k-1})$, $i=1,\dots,N-1$, is applied over the complete prediction horizon (except the last step). Conceptually, the sequence of delta sets can be interpreted as a tube $\Delta^{N-1}(\mathcal{X}^\star_{k-1}):=(\Delta(\bm{x}^\star_{2|k-1}),\dots,\Delta(\bm{x}^\star_{N|k-1}))$ that constrains the predicted state sequence (which will make sure in the end that we have constraint satisfaction, as we will prove in Theorem~\ref{lem:rec_feas_terminal_set}). Note that the role of the tube is different from tube-based MPC \cite{Mayne2005}. In Fig.~\ref{fig:ca-MPC_illustration_terminal} we provide an illustrative example of the ca-MPC scheme from \eqref{eq:reduced_MPC_terminal}.
\begin{figure}[!ht]
	\centering
	\begin{tikzpicture}[join=round]
		\definecolor{red}{HTML}{eb2f06};
		\definecolor{blue}{HTML}{2980b9};
		\definecolor{green}{HTML}{44bd32};
		
		%	Draw constraint set (counter clockwise starting from left vertical)
		\draw[very thick] (0.8, 1.8)--(0.4, 3.8);
		\draw[very thick] (0.3, 2.2)--(5.6, 1.9);
		\draw[very thick] (5.1, 1.7)--(7, 3.5);
		\draw[very thick] (7, 3)--(5.5, 5);
		\draw[very thick, color = red] (6.3, 4.7)--(2.7, 5);
		\draw[very thick, color = red] (4.5, 5)--(1, 4.0);
		\draw[very thick, color = red] (3.5, 5.2)--(0.1, 3.2);
		
		%	Draw terminal set
		\draw[very thick,color=blue] (4.8,3.9) -- (6.2,3.5) -- (5,2.3) -- (4.0,2.7) -- cycle;
		
		%	Draw state trajectory
		%coordinates
		\coordinate (x0) at (1.3,2.8);
		\coordinate (x1) at (2.2,3.9);
		\coordinate (x2) at (3.4,4.4);
		\coordinate (x3) at (5.0,4.78);
		\coordinate (x4) at (5.7,3.4);	
		
		\draw[very thick, dashed]  (x0) -- (x1) -- (x2) -- (x3) -- (x4) ;
		
		\draw[very thick, color=green] (x2) 	circle(1.15);
		\node[color=green] at ($ (x2) + (-1.7,0.9) $) {$\Delta(\bm{x}^\star_{2|k-1})$};
		\draw[very thick, color=green] (x3) 	circle(0.7);
		\node[color=green] at ($ (x3) + (1.5,0.5) $) {$\Delta(\bm{x}^\star_{3|k-1})$};
		\draw[very thick, color=green] (x4)		circle(0.75);
		\node[color=green] at ($ (x4) + (1.25,0.8) $) {$\Delta(\bm{x}^\star_{4|k-1})$};
				
		\draw[fill=black] (x0) circle(0.05) node[below right] {$\bm{x}^\star_{0|k-1}$};
		\draw[fill=black] (x1) circle(0.05) node[below right] {$\bm{x}^\star_{1|k-1}$};
		\draw[fill=black] (x2) circle(0.05) node[below right] {$\bm{x}^\star_{2|k-1}$};
		\draw[fill=black] (x3) circle(0.05) node[above] {$\bm{x}^\star_{3|k-1}$};
		\draw[fill=black] (x4) circle(0.05) node[below] {$\bm{x}^\star_{4|k-1}$};
			
		%	Annotate sets
		\node[] at (3.4,2.3) {$\mathbb{X}$};
		\node[color=blue] at (4.6,2.8) {$\mathbb{X}_T$};
		\node[color=red] at (1.25,4.55) {$\mathbb{X}_r(\mathcal{X}^\star_{k-1})$};
	\end{tikzpicture}
	\vspace{-0.3cm}
	\caption{Illustrating ca-MPC with a terminal set and parameterized by $\mathcal{X}^\star_{k-1}$. Note that $\mathbb{X}_r(\mathcal{X}^\star_{k-1})$ is constructed as detailed in Remark~\ref{rem:invariant_reduced_constraint_set} for illustration purposes. Critically, the reduced constraint set and delta sets ensure $\bm{x}_{i|k}\in\mathbb{X}$ for $i=1,\dots,N$.}
	\label{fig:ca-MPC_illustration_terminal}
\end{figure}
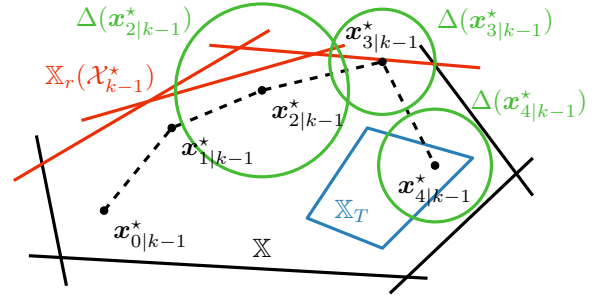
\begin{rem}
We assume the terminal set is sufficiently simple to be included in \eqref{eq:reduced_MPC_terminal} without computational issues. We note that reduction of the terminal set constraints is possible as well, but it is not detailed for ease of exposition.
\end{rem} 

In particular, the constraints in Problem~\eqref{eq:reduced_MPC_terminal} are well-defined (and parameterized) for all $\mathcal{X}_{k-1} =\\ \allowbreak (\bm{x}_{1|k-1},\dots,\bm{x}_{N|k-1})$ contained in the 
 set $\mathbb{S}$ defined as 
\begin{align} \label{eq:S}
\mathbb{S}:= \{(\bm{\tilde x}_{1},\dots,\bm{\tilde x}_{N})\in \mathbb{X}^{N} \mid \text{ there exists } \notag \\ {(\bm{\tilde u}_{1},\dots,\bm{\tilde u}_{N-1})\in \mathbb{U}^{N-1}} \text{ s.t. } \bm{\tilde x}_{i+1} = f(\bm{\tilde x}_{i},\bm{\tilde u}_{i}), \notag\\ \ i=1,\ldots, N-1, \text{ and } \bm{\tilde x}_{N} \in \mathbb{X}_T \}, 
\end{align}

which are all state trajectories of length $N$ that satisfy the system dynamics \eqref{eq:system_description}, the input and state constraints \eqref{eq:state_and_input_constraints}, and end in the terminal set $\mathbb{X}_T$. 

For $k=0$, the sets $\mathbb{X}_r(\bm{x}_{i+1|k-1})$ for $i=1,\dots,N-1$ are not explicitly defined yet as the state sequence $\mathcal{X}^\star_{-1}$ is not available. However, to overcome this, the scheme is started up at $k=0$ and state $\bm{x}_0$ by finding an $\mathcal{X}_{-1}^\star \in \mathbb{S}$, where $\mathbb{S}$ as in \eqref{eq:S} with the additional constraint that $\bm{\tilde{x}}_1=\bm{x}_0$. With such a sequence $\mathcal{X}_{-1}^\star\in\mathbb{S}$ the optimization problem \eqref{eq:reduced_MPC_terminal} can be formulated and solved, and this can be considered as the initial condition to our ca-MPC scheme. Therefore, we introduce 
\begin{multline}
\mathbb{S}_0(\bm{x}_0):= \{(\bm{\tilde x}_{1},\bm{\tilde x}_{2},\dots,\bm{\tilde x}_{N})\in \mathbb{S} \mid \bm{\tilde x}_{1}=\bm{x}_0 \}.
\end{multline}

We denote the set of feasible input sequences to \eqref{eq:reduced_MPC_terminal} given $\mathcal{X}_{k-1}^\star \in \mathbb{S}$, by $\mathbb{U}_F^T(\mathcal{X}_{k-1}^\star) :=\{ (\bm{u}_{0|k},\dots,\bm{u}_{N-1|k}) \in \mathbb{U}^N \mid \eqref{eq:regular_MPC_terminal_b},\eqref{eq:regular_MPC_terminal_c},\eqref{eq:regular_MPC_terminal_e},\eqref{eq:regular_MPC_terminal_f},\eqref{eq:reduced_MPC_terminal_d}\}$, and corresponding state trajectories by 
\begin{align}
\mathbb{X}_F^T(\mathcal{X}_{k-1}^\star) :=\{ &(\bm{x}_{1|k},\dots,\bm{x}_{N|k}) \in (\mathbb{R}^n)^{N} \mid \eqref{eq:regular_MPC_terminal_b},\eqref{eq:regular_MPC_terminal_c}\notag\\ \text{ for some } &(\bm{u}_{0|k},\dots,\bm{u}_{N-1|k}) \in \mathbb{U}_F^T(\mathcal{X}^\star_{k-1}) \}.
\end{align}
The resulting MPC law based on the receding-horizon principle is denoted by $K_{MPC}^{T,r}:\mathbb{S}\rightarrow \mathbb{U}$, i.e., 
\begin{align} \label{eq:kmpctr}
\bm{u}_k = K_{MPC}^{T,r}(\mathcal{X}_{k-1}^\star) = \bm{u}^\star_{0|k}, \ k \in \mathbb{N}
\end{align}
where 
$\mathcal{U}^\star_{k} := (\bm{u}^\star_{0|k},\dots,\bm{u}^\star_{N-1|k})$ is a minimizer in \eqref{eq:reduced_MPC_terminal}, i.e., 
\begin{align}
\mathcal{U}^\star_{k} \in \arg \min_{\mathcal{U}_k \in \mathbb{U}_F^T(\mathcal{X}_{k-1}^\star) } \ell_T(\bm{x}_{N|k}) + \sum_{i=0}^{N-1}\ell(\bm{x}_{i|k},\bm{u}_{i|k}).
\end{align}
For $k=0$, we select $ \mathcal{X}_{-1}^\star \in \mathbb{S}_0(\bm{x}_0)\subset \mathbb{S}$. Note that $\bm{u}_k$ in this MPC scheme depends on $\mathcal{X}_{k-1}^\star$, which includes $\bm{x}_k$, but thus also depends on the future predicted states at the previous time step. 

Similar to the previous section, recursive feasibility and constraint satisfaction are guaranteed by imposing suitable conditions on $\Delta^{N-1}(\mathcal{X})$.

\begin{thm}\label{lem:rec_feas_terminal_set}
	Consider system \eqref{eq:system_description} with the reduced MPC scheme \eqref{eq:reduced_mpc_scheme} and let the terminal set $\mathbb{X}_T\subseteq \mathbb{X}$ satisfy Assumption~\ref{ass:target_set}. Suppose that for all $\mathcal{X}=(\bm{\tilde x}_{1},\dots,\bm{\tilde x}_{N}) \in \mathbb{S}$, the tube $\Delta^{N-1}(\mathcal{X})$ satisfies
	\begin{enumerate}
		\item $\mathbb{X}_r( \bm{\tilde x}_{i})\cap\Delta(\bm{\tilde x}_{i}) \subseteq \mathbb{X}$ for $i=2,3,\dots,N$,\label{cond:delta_cond3}
		\item $\bm{\tilde x}_{i} \in \Delta(\bm{\tilde x}_{i}) $ for $i=2,3,\dots,N$. \label{cond:delta_cond4}
	\end{enumerate}
	Then, the reduced MPC scheme \eqref{eq:reduced_MPC_terminal} satisfies the following properties:\
	\begin{description}
	\item[Recursive feasibility]$~$\\ Given $\mathcal{X}_{k-1} \in \mathbb{S}$, then $\mathbb{U}_F^T(\mathcal{X}_{k-1})\allowbreak\neq\allowbreak\emptyset$ and any $\mathcal{X}_{k}\in \mathbb{X}_F^T(\mathcal{X}_{k-1})$ (including any optimal $\mathcal{X}_{k}^\star\in \mathbb{X}_F^T(\mathcal{X}_{k-1})$) satisfies $\mathcal{X}_k\in \mathbb{S}$. 
	\item[Constraint satisfaction]$~$\\ For all $\bm{x}_0\in\mathbb{X}$ with $\mathcal{X}^\star_{-1} \in \mathbb{S}_0(\bm{x}_0)$, the trajectories of the closed-loop system \begin{equation}
	\bm{x}_{k+1} = f(\bm{x}_k,\bm{u}_k),\ \bm{u}_k= K_{MPC}^{T,r}(\mathcal{X}_{k-1}^\star) 
	\end{equation} are well-defined for all $k\in \nat$
	and the constraints in \eqref{eq:state_and_input_constraints} are satisfied for all $k\in{\mathbb N}$. 
	\end{description} 
\end{thm}

We now have sufficient conditions on the structure of $\Delta^{N-1}(\cdot)$ that guarantee recursive feasibility and constraint satisfaction. Again, an example for a simple tube $\Delta^{N-1}(\mathcal{X}_{k-1})$ that satisfies Conditions~\eqref{cond:delta_cond3} and~\eqref{cond:delta_cond4} given $\mathcal{X}_{k-1} \in \mathbb{S}$, is a sphere centered around $\mathcal{X}_{k-1}$: $\Delta(\bm{x}_{i+1|k-1}) = \{\bm{x}\in\mathbb{R}^n\mid \|\bm{x}_{i+1|k-1} - \bm{x}\|_2 \leq c(\bm{x}_{i+1|k-1})\}$. Here, we increase the radius $c(\bm{x}_{i+1|k-1})$ subject to $\mathbb{X}_r(\bm{x}_{i+1|k-1})\cap\Delta(\bm{x}_{i+1|k-1}) \subseteq \mathbb{X}$. Observe that the conditions in Theorem~\ref{lem:rec_feas_terminal_set} are trivially satisfied by these choices and that $c(\bm{x}_{i+1|k-1})=0$ is always a valid choice, although, clearly, large values of $c(\bm{x}_{i+1|k-1})$ are desirable as
it gives more freedom in the MPC optimization problem. This particular choice of $\Delta^{N-1}(\mathcal{X}_{k-1})$ is illustrated in Fig.~\ref{fig:ca-MPC_illustration_terminal}.
\begin{rem}\label{rem:invariant_reduced_constraint_set}
	If a reduced constraint set that varies over the horizon is undesired, it is also possible to replace $\mathbb{X}_r(\bm{x}_{i+1|k-1})$ by $\mathbb{X}_r({\mathcal{X}_{k-1}}):= \bigcap_{i=2}^{N} \mathbb{X}_r(\bm{x}_{i|k-1}).$
\end{rem} 
\begin{rem} \label{rem:stability}
Interestingly, it can be established that if the terminal set and cost are chosen to guarantee asymptotic stability of the origin for the closed-loop system under the common MPC setup \eqref{eq:regular_MPC_terminal} using the typical terminal conditions, see, e.g. (A1-A4) in \cite{Mayne2000}, then this also guarantees asymptotic stability of the origin under the ca-MPC control strategy \eqref{eq:reduced_MPC_terminal}. 
\end{rem}

%%%%%%%%%%%%%%%%%%%%%%%%%%%%%%%%%%%%%%%

\section{Double integrator case study}\label{sec:sim_results}
To demonstrate ca-MPC we apply the proposed methodology in Section~\ref{sec:terminal} to a double integrator system. We opted to use this two-dimensional system as it is more easy to visualize and we can, just as for large-scale systems, impose many (non-redundant) state constraints, which makes this a good example for illustration. Additionally, the non-reduced MPC solution \eqref{eq:regular_MPC_terminal} is tractable for this system, which allows for a comparison to our ca-MPC scheme. Finally, our implementation uses the inputs as decision variables (so-called ``dense'' formulation) which makes the model order less relevant. Hence, the double integrator system with many state constraints is a good representative for more complex systems. 
\subsection{System and constraint definition}
The double integrator system is described by
\begin{align}\label{eq:double_integrator}
	\bm{x}_{k+1} = \begin{bmatrix}
		1 & 0.1\\ 0 & 1
	\end{bmatrix} \bm{x}_k + \begin{bmatrix}
		0.005 \\ 0.1
	\end{bmatrix}u_k.
\end{align}
Next, the state constraints are defined as piecewise linear approximations of two quadratic constraints resulting in $n_x = 1000$ constraints, see Fig.~\ref{fig:mpc_solution}. The input is constrained by $\mathbb{U} = [-0.5,0.5]$ and the prediction horizon is chosen as $N=15$. Last, the terminal set is shown in Fig.~\ref{fig:mpc_solution}. The resulting quadratic program given by \eqref{eq:regular_MPC_terminal} has approximately $14,000$ linear inequality constraints.

We implement the ca-MPC scheme from Section~\ref{sec:terminal} as $\mathbb{X}$ is not controlled invariant. Here, we choose $\ell(\bm{x},u) = \|\bm{x}\|_2^2 + u^2 $ and $\ell_T(\bm{x}) = \|\bm{x}\|_2^2 $. Next, $\bm{\alpha}$, which defines the safety set and in turn the reduced constraint set, is given by
\begin{align}
	\bm{\alpha} = 0.1 \begin{bmatrix}\|\bm{C}_{1,:}\|_2 & \dots & \|\bm{C}_{n_x,:}\|_2	\end{bmatrix}^\top.
\end{align}
This has the interpretation that $\mathbb{X}$ is tightened by a uniform distance. Recall that $\bm{C} = \begin{bmatrix}\bm{C}_{1,:}^\top &\dots&\bm{C}_{n_x,:}^\top \end{bmatrix}^\top$ is the matrix defining the state constraints \eqref{eq:state_and_input_constraints}. Additionally, for ease of implementation, we use a reduced constraint set that does not vary over the horizon, see Remark~\ref{rem:invariant_reduced_constraint_set}. Next, we define the delta set as $\Delta(\bm{x}) := \{\bm{\hat{x}}\in\mathbb{R}^2\mid \|\bm{\hat{x}}-\bm{x}\|_\infty \leq c(\bm{x})\}$. The radius $c(\bm{x})$ is chosen such that $\Delta(\bm{x}_{i+1|k-1}) \subseteq \mathbb{X}_c(\mathcal{X}_k)$, where
\begin{subequations}
 \begin{align}
	& \mathbb{X}_c(\mathcal{X}_k):=\{\bm{x}\in\mathbb{R}^2\mid \bm{C}_j\bm{x}\leq b_j,\ j\in\mathcal{J}_c(\mathcal{X}_k)\},\\
	& \mathcal{J}_c(\mathcal{X}_k) := \mathbb{N}_{[1,n_x]}\backslash\ \left(\bigcap_{i=2}^N \mathcal{J}(\bm{x}_{i|k})\right).
\end{align}  
\end{subequations}
Observe that $\mathbb{X}_c(\mathcal{X}_k)$ contains the constraints not included in $\mathbb{X}_r(\mathcal{X}_k)$. The radius $c(\bm{x})$ is chosen by checking the vertices of $\Delta(\bm{x})$ against $\mathbb{X}_c(\mathcal{X})$. Note that this choice for $\mathbb{X}_c(\mathcal{X}_k)$ is conservative in the light of Condition~\eqref{cond:delta_cond3} from Theorem~\ref{lem:rec_feas_terminal_set}. However, the additional conservatism is considered acceptable given the resulting trivial construction of the delta sets. Of course, all guarantees regarding constraint satisfaction and recursive feasibility remain to hold.

\subsection{Results}
\begin{figure}[!ht]
	\centering
	\includegraphics[width=8cm]{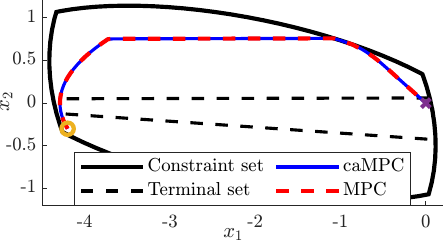}
	 \vspace{-0.3cm} 
	\caption{The closed-loop state trajectory using the regular MPC solution and the ca-MPC solution.}
	\label{fig:mpc_solution}
\end{figure}

\begin{figure}[!ht]
	\centering
	\includegraphics[width=9cm]{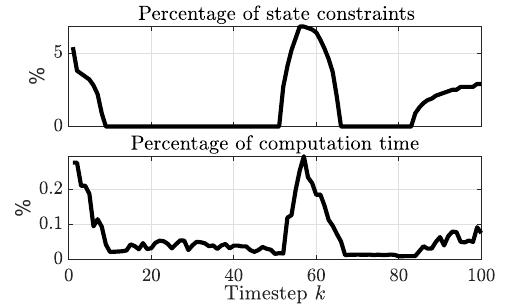}
	\vspace{-0.7cm}
	\caption{The percentage of constraints and the computation time with respect to the full MPC solution over time. When the state approaches the boundary of $\mathbb{X}$ the number of constraints and computation time increase.} % Nevertheless, significant speed improvements are observed.}
	\label{fig:ca-MPC_speedup}
\end{figure}

The ca-MPC closed-loop trajectory is compared with the (unreduced) MPC solution from \eqref{eq:basic_mpc}, see Fig.~\ref{fig:mpc_solution}. Here, \eqref{eq:double_integrator} is initialized at $\bm{x}_0 = \begin{bmatrix}-4.2& -0.3\end{bmatrix}^\top$ and steered towards the origin. As seen in Fig.~\ref{fig:mpc_solution}, the ca-MPC solution is indistinguishable from the MPC solution. Nevertheless, ca-MPC uses only a fraction of the state constraints and has significantly lower computation time. The percentage of state constraints and the computational time, both with respect to traditional MPC, are shown in Fig.~\ref{fig:ca-MPC_speedup}. Note that the computation of the delta sets and reduced constraint set is included in the computation time for ca-MPC. From Fig.~\ref{fig:ca-MPC_speedup} we observe an improvement in the computation time of two orders of magnitude. This result clearly demonstrates the potential of ca-MPC for systems with many state constraints.
\vspace{-0.3cm}

%%%%%%%%%%%%%%%%%%%%%%%%%%%%%%%%%%%%%%

\section{Conclusion}\label{sec:conclusion}
In this paper, we presented \textit{constraint-adaptive} MPC (ca-MPC) schemes that enable real-time feasibility for complex nonlinear systems with a large number of state inequality constraint. The two novel ca-MPC designs were based on the idea of dynamically varying the set of imposed state constraints in the MPC optimization problem in such a way that recursive feasibility and constraint satisfaction is guaranteed. The first scheme exploited the controlled invariance of the original state constraint set, while the second relaxed this assumption and just required the availability of a basic terminal set. A simulation study with a benchmark double integrator system showed the potential of our new framework. Indeed, the implemented ca-MPC lead to a significant reduction of the computation time with a negligible impact on closed-loop performance.

%%%%%%%%%%%%%%%%%%%%%%%%%%%%%%%%%%%%%%%%

This work can be extended in various directions. For instance, as indicated in Remark~\ref{rem:stability}, it is possible to preserve stability of the closed-loop ca-MPC system using standard conditions on the terminal ingredients. Interesting future work in this light is to extend this preservation of closed-loop stability also to performance guarantees by proper construction of safety and delta sets. Also the extension of our results to ca-MPC setups with large prediction horizons, without the use of controlled invariance of the state constraint set or the use terminal sets, is of interest.

%%%%%%%%%%%%%%%%%%%%%%%%%

\bibliographystyle{IEEEtran}
\bibliography{ifacconf} % bib file to produce the bibliography

\begin{thebibliography}{10}
\providecommand{\url}[1]{#1}
\csname url@samestyle\endcsname
\providecommand{\newblock}{\relax}
\providecommand{\bibinfo}[2]{#2}
\providecommand{\BIBentrySTDinterwordspacing}{\spaceskip=0pt\relax}
\providecommand{\BIBentryALTinterwordstretchfactor}{4}
\providecommand{\BIBentryALTinterwordspacing}{\spaceskip=\fontdimen2\font plus
\BIBentryALTinterwordstretchfactor\fontdimen3\font minus
  \fontdimen4\font\relax}
\providecommand{\BIBforeignlanguage}[2]{{%
\expandafter\ifx\csname l@#1\endcsname\relax
\typeout{** WARNING: IEEEtran.bst: No hyphenation pattern has been}%
\typeout{** loaded for the language `#1'. Using the pattern for}%
\typeout{** the default language instead.}%
\else
\language=\csname l@#1\endcsname
\fi
#2}}
\providecommand{\BIBdecl}{\relax}
\BIBdecl

\bibitem{Mayne2014}
D.~Q. Mayne, ``{Model predictive control: Recent developments and future
  promise},'' \emph{Automatica}, vol.~50, no.~12, pp. 2967--2986, 2014.

\bibitem{Dotoli2015}
M.~Dotoli, A.~Fay, M.~Mi{\'{s}}kowicz, and C.~Seatzu, ``{A Survey on Advanced
  Control Approaches in Factory Automation},'' \emph{IFAC-PapersOnLine},
  vol.~48, no.~3, pp. 394--399, 2015.

\bibitem{Rawlings2019}
J.~B. Rawlings, D.~Q. Mayne, and M.~M. Diehl, \emph{{Model Predictive Control:
  Theory, Computation, and Design}}, 2nd~ed.\hskip 1em plus 0.5em minus
  0.4em\relax Nob Hill Publishing, 2019.

\bibitem{Hovland2006}
S.~Hovland, K.~Willcox, and J.~T. Gravdahl, ``{MPC for Large-Scale Systems via
  Model Reduction and Multiparametric Quadratic Programming},'' in \emph{45th
  IEEE Conf. on Decision and Control}, 2006, pp. 3418--3423.

\bibitem{Jerez2011}
J.~L. Jerez, E.~C. Kerrigan, and G.~A. Constantinides, ``{A condensed and
  sparse QP formulation for predictive control},'' in \emph{50th IEEE Conf. on
  Decision and Control and European Control Conf.}, dec 2011, pp. 5217--5222.

\bibitem{Boyd2009}
S.~Boyd and L.~Vandenberghe, \emph{{Convex optimization}}, 7th~ed.\hskip 1em
  plus 0.5em minus 0.4em\relax Cambridge: Cambridge University Press, 2009.

\bibitem{Altmuller2012}
N.~Altm{\"{u}}ller and L.~Gr{\"{u}}ne, ``{Distributed and boundary model
  predictive control for the heat equation},'' \emph{GAMM-Mitteilungen},
  vol.~35, no.~2, pp. 131--145, nov 2012.

\bibitem{Hendrikx2018}
R.~Hendrikx, S.~Curto, B.~De~Jager, E.~Maljaars, G.~Van~Rhoon, M.~Paulides, and
  W.~Heemels, ``Pod-based recursive temperature estimation for mr-guided rf
  hyperthermia cancer treatment: A pilot study,'' in \emph{IEEE Conf. on
  Decision and Control}, 2018, pp. 5201--5208.

\bibitem{Paulides2013}
M.~M. Paulides, P.~R. Stauffer, E.~Neufeld, P.~F. Maccarini, A.~Kyriakou, R.~A.
  Canters, C.~J. Diederich, J.~F. Bakker, and G.~C. {Van Rhoon}, ``{Simulation
  techniques in hyperthermia treatment planning},'' \emph{International Journal
  of Hyperthermia}, vol.~29, no.~4, pp. 346--357, jun 2013.

\bibitem{Antoulas2000}
S.~{Antoulas, A.C.; Sorensen, D.C.; Gugercin}, ``{A Survey of Model Reduction
  Methods for Large-Scale Systems.}'' Tech. Rep., 2000.

\bibitem{Paulraj2010}
S.~Paulraj and P.~Sumathi, ``{A comparative study of redundant constraints
  identification methods in linear programming problems},'' \emph{Mathematical
  Problems in Engineering}, vol. 2010, 2010.

\bibitem{Jost2013}
M.~Jost and M.~Mönnigmann, ``{Accelerating model predictive control by online
  constraint removal},'' in \emph{52nd IEEE Conf. on Decision and Control}, dec
  2013, pp. 5764--5769.

\bibitem{Hertneck2018}
M.~Hertneck, J.~Köhler, S.~Trimpe, and F.~Allgöwer, ``{Learning an
  Approximate Model Predictive Controller With Guarantees},'' \emph{IEEE
  Control Systems Letters}, vol.~2, no.~3, pp. 543--548, jul 2018.

\bibitem{Klauco2019}
M.~Klau{\v{c}}o, M.~Kal{\'{u}}z, and M.~Kvasnica, ``{Machine learning-based
  warm starting of active set methods in embedded model predictive control},''
  \emph{Engineering Applications of Artificial Intelligence}, vol.~77, pp.
  1--8, jan 2019.

\bibitem{Mayne2000}
D.~Q. Mayne, J.~Rawlings, C.~Rao, and P.~Scokaert, ``{Constrained model
  predictive control: Stability and optimality},'' \emph{Automatica}, vol.~36,
  no.~6, pp. 789--814, jun 2000.

\bibitem{Mayne2005}
D.~Q. Mayne, M.~Seron, and S.~Rakovi{\'{c}}, ``{Robust model predictive control
  of constrained linear systems with bounded disturbances},''
  \emph{Automatica}, vol.~41, no.~2, pp. 219--224, feb 2005.

\end{thebibliography}
  % with bibtex (preferred)

\end{document}